\documentclass[a4paper,12pt]{article}
\usepackage{amsfonts,amsthm,amsmath,amssymb,graphicx,hyperref,color,youngtab}

\newcommand{\arXiv}[2]{\href{http://arxiv.org/abs/#1}{{\tt arXiv:#2}}}
\newcommand{\hep}[2]{\href{http://arxiv.org/abs/#1}{{\tt #2}}}

\newcommand{\tr}{\operatorname{tr}}

\newcommand{\be}{\begin{equation}}
\newcommand{\bea}{\begin{eqnarray}}

\newcommand{\ee}{\end{equation}}
\newcommand{\eea}{\end{eqnarray}}

\begin{document} 

\makeatletter
\@addtoreset{equation}{section}
\makeatother
\renewcommand{\theequation}{\thesection.\arabic{equation}}

\vspace{1.8truecm}

\vspace{15pt}


{\LARGE{ 
\centerline{\bf Structure Constants of a Single Trace Operator and  } 
\centerline{\bf Determinant Operators from Hexagon } 
}}  

\vskip.7cm 

\thispagestyle{empty} 
\centerline{ {\large\bf Keun-Young Kim$^{a}$\footnote{{\tt fortoe@gist.ac.kr}}, 
Minkyoo Kim$^{b,}$\footnote{{\tt mimkim80@gmail.com}} }}

\centerline{ {\large\bf and Kyungsun Lee${}^{a,}$\footnote{ {\tt kyungsun.cogito.lee@gmail.com}}}}

\vspace{.5cm}
\centerline{{\it ${}^a$ School of Physics and Chemistry},}
\centerline{{ \it Gwangju Institute of Science and Technology, Gwangju 61005,
Korea}}

\vspace{.4cm}
\centerline{{\it ${}^b$ National Institute for Theoretical Physics,}}
\centerline{{\it School of Physics and Mandelstam Institute for Theoretical Physics,}}
\centerline{{\it University of the Witwatersrand, Wits, 2050, } }
\centerline{{\it South Africa } }

\vspace{1truecm}

\thispagestyle{empty}

\centerline{\bf ABSTRACT}

\vskip.2cm 

We study the structure constant of a single trace operator and two determinant operators in ${\cal N}=4$ super Yang-Mills theory. 
Holographically such a quantity corresponds to the interaction vertex between a closed string and two open strings attached to the spherical $D$-branes. 
Relying on diagrammatic intuition, we conjecture that the structure constant at the finite coupling is nicely written by the hexagon form factors.
Precisely we need to prepare two hexagon twist operators and appropriately glue edges together by integrating mirror particles contributions and by contracting boundary states. 
The gluing generates the worldsheet for a closed string and two open strings attached to the $D$-branes.
At the weak coupling, the asymptotic expression simply reduces to sum over all possible partitions not only for the edge related to the closed string but also for the edges representing the half of the open string together with reflection effects for the opposite open string edges. 
We test the conjecture by directly computing various tree level structure constants. 
The result is nicely matched with our conjecture.

\setcounter{page}{0}
\setcounter{tocdepth}{2}
\newpage
\tableofcontents
\setcounter{footnote}{0}
\linespread{1.1}
\parskip 4pt

{}~
{}~

\section{Introduction}

The fusion of holographic duality and quantum integrability has led to numerous surprising results in computing nonperturbatively physical observables of both string theory and gauge field theory in the planar limit \cite{Beisert:2010jr}. 
It is within bounds to say that the core idea in these developments is the exact $S$-matrix \cite{Beisert:2005tm, Janik:2006dc, Beisert:2006ez, Arutyunov:2006yd}.
This is because one can unify theoretical concepts such as spin-chain\cite{Staudacher:2004tk}, light-cone gauge string theory\cite{Klose:2006zd}, all-loop asymptotic Bethe ansatz\cite{Beisert:2005fw} and finite size corrections\cite{Janik:2007wt, Bajnok:2008bm} in terms of the exact $S$-matrix.
Even though the matrix part of $SU(2|2)$ $S$-matrix by Beisert \cite{Beisert:2005tm} was derived only by symmetry argument rather than a bootstrap idea or integrability condition represented through Yang-Baxter equation, completely determining the scalar phase was essential for computing physical data at the finite coupling.
In the problem of the scalar part, the bootstrap idea such as unitarity and crossing symmetry was necessary \cite{Janik:2006dc}. 
Furthermore, it was shown that the full $S$-matrix satisfies Yang-Baxter equation which is the most transparent condition of quantum integrability \cite{Arutyunov:2006yd}.
The spectral problem of the planar ${\cal N}=4$ SYM has finally solved in the quantum spectral curve which is nothing but the most developed formalism based on the exact $S$-matrix \cite{Gromov:2014caa}.

Beyond the spectral problem, the form factor idea known well in integrable models was started to apply in AdS/CFT correspondence \cite{Bajnok:2015hla, BKV}.
Basically, the spectral problem is about the two-point function of conformal gauge theory. On the other hand, with the form factor idea, one can treat three-point and higher correlation functions.
Although this development is still on-going, fascinating results in various contexts and setups were already proposed and tested \cite{Bajnok:2015ftj, Bajnok:2017mdf, Jiang:2015lda, Eden:2015ija, Basso:2015eqa, Eden:2016xvg, Fleury:2016ykk, Basso:2017muf, Kim:2017phs, Kim:2017sju, Bargheer:2017nne, Fleury:2017eph, Eden:2018vug, Bargheer:2018jvq, Nieto:2018izd, Kiryu:2018phb, Basso:2018cvy, Kostov:2019stn, Bargheer:2019kxb, Kostov:2019auq}. 
Unlike the form factor in integrable relativistic quantum field theories, the form factor approaches used in computing correlation functions of the ${\cal N}=4$ SYM is usually defined by not a local operator but a twist operator. 
The twist operator has some branch points and sometimes can have multiple in and out states \cite{Cardy:2007mb}.
Among them, the most promising and studied example is to consider hexagon twist operators in correlator problem. 

How can we figure out the hexagon intuitively?
Interestingly, the starting point of strategy computing correlation functions through the hexagon twist operator is input from an intuitive diagram.  
Surely, we know that intuition through pictorial description does not always work. 
Nevertheless, we should also admit that diagrammatic representation is quite helpful in many physical theories or sometimes quite essential. Think of electric flux lines, image charge method, Feynman diagram, etc.
If we cut so-called the pair of pants diagram of interacting three closed strings(or spin-chains) into two pieces along three seams, we would get two hexagons \cite{BKV}. Interestingly, with the hexagon which is nothing but the half of worldsheet, one can get exact form factor which is valid at finite coupling.\footnote{For pedagogical review, see \cite{Les}.}
Actually, the hexagonalization idea is to reconstruct the whole worldsheet in terms of hexagon.\footnote{To obtain final results, we need to consider all mirror particle contributions. Nevertheless, the product of hexagon with summing over all possible partitions is still valid asymptotically.} 
For example, the hexagonialization of four-point correlator is different with the OPE decomposition through merging between two three-point vertices \cite{Fleury:2016ykk}.
The clear difference can be also shown graphically without any complicated explanation.
Nevertheless, we would not know a priori if such hexagons indeed satisfy mathematical consistency and match with data obtained by the brute force method. 
Namely, we should do a sort of experiments to check the proposal. Till now, perturbative calculations at the weak coupling regime really seem to work since those were consistent with the hexagon proposal.
At the strong coupling regime, correlation functions of strings were not fully reproduced by the hexagon. Nevertheless, very nontrivial part of full results was tested well.

The hexagon idea was also generalized to the vertex of three open strings \cite{Kim:2017phs, Kiryu:2018phb}.
In this case, there is no cutting since we need just one hexagon twist operator. On the other hand, it was noticed that gluing boundary states into hexagon is necessary \cite{Kim:2017phs, Kiryu:2018phb}.
In the previous works about three-open string structure constant in terms of hexagon, the set-up is realized by small deformation of $1/2$-BPS Wilson loop in dual gauge theory. 
On the other hand, there exist some other integrability preserving boundary conditions which are quite different from the Wilson loop problem. 
One of them is the open string attached to the giant graviton brane \cite{Berenstein:2005vf, Hofman:2007xp}. 
We may expect that such a three-point function of open strings attached to the giant graviton can be similarly studied by hexagon idea.
However, we found that in the three open strings problem, it is not possible to have a simple set-up where $D$-brane part and open string part are decoupled each other.
Actually, the simplest set-up for studying integrable open strings attached to the giant graviton in hexagon idea turns out to be the problem of two open strings and a closed string.\footnote{Three-point function for an open string and two closed string is also similarly excluded since there is nontrivial mixing between $D$-brane part and open string part.}

\begin{figure}[t]
\begin{center}
\includegraphics[trim={5cm 9cm 10cm 3cm},clip, width=8cm]{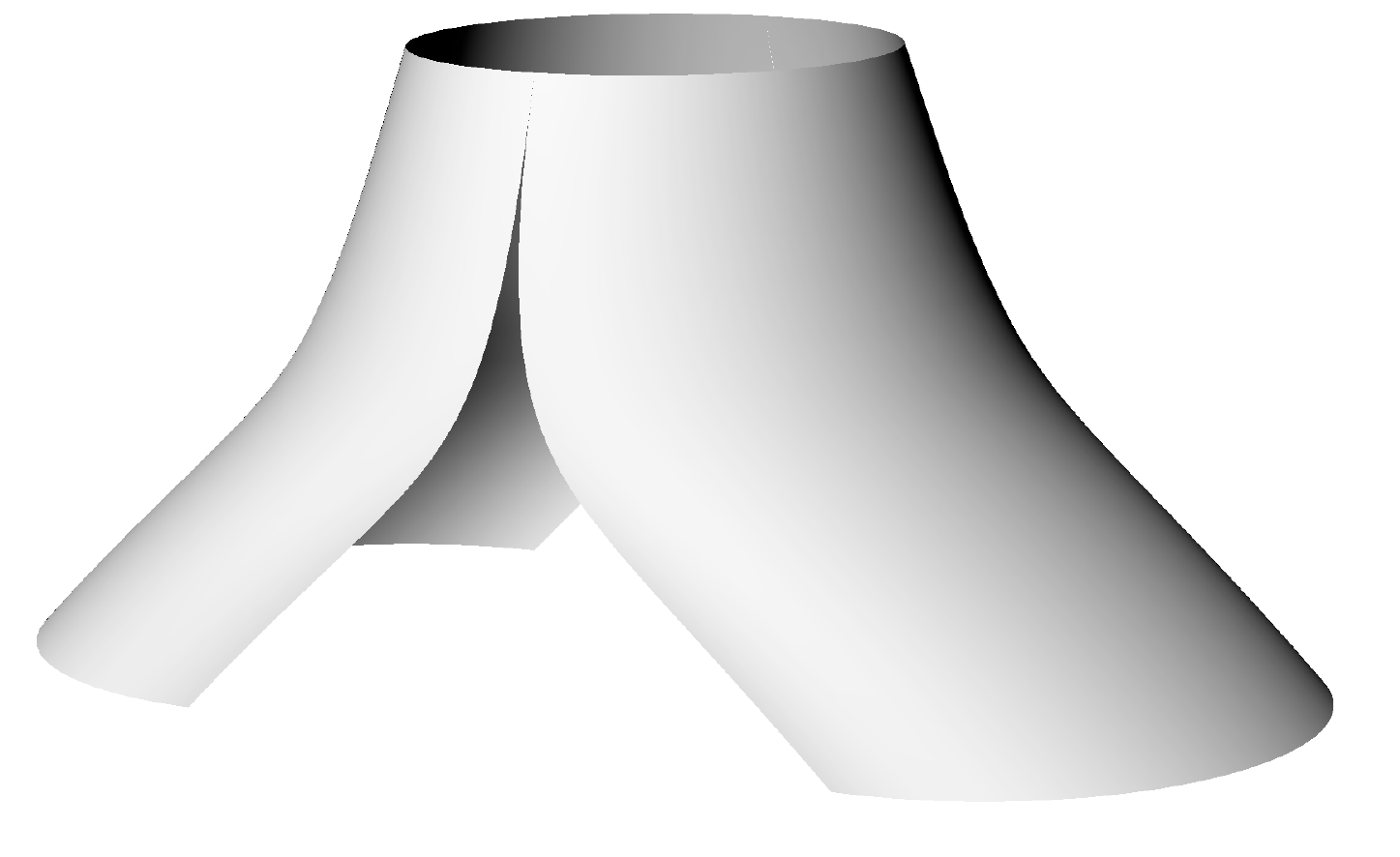}
\caption{Worldsheet diagram for merging two open strings into a closed string}
\label{wsd}
\end{center}
\end{figure}

Diagrammatic description for two open strings attached to the giant graviton brane and a closed string is given in Figure \ref{wsd}.
Namely, two open strings move towards each other, form a closed string by merging together and the closed string propagates apart from $D$-brane to the bulk.
Can we conjecture some meaningful results for this physical situation in terms of hexagons?
Our answer is yes. In this paper, we shall give a conjecture for the finite coupling result and test if the conjecture makes sense at the weak coupling regime.

In section 2, we shall introduce our set-up and conventions.
In section 3, we shall make a finite coupling conjecture for structure constant of two open strings attached to the giant graviton brane and a closed string. 
In section 4, we shall compute the structure constants of two determinant operators and a single trace operator at the weak coupling tree level. In particular, we shall choose the simplest setup where is no contraction between the $D$-brane part and the closed string part.
The contractions between two determinants and between a determinant and a single trace operator would be performed only through strings.
Finally, We finish this paper with a discussion. 
We also give an appendix for the  integrable open spin-chain used in this paper.

\noindent{\bf Note added:} While we were writing up this article, a relevant paper appeared on arxiv \cite{Jiang:2019xdz}. 
The paper suggested interesting new techniques and treated interactions between $D$-brane part and closed string part which we are ignoring in this paper.
Concretely, they studied structure constants for two determinants and a non-BPS single trace operators where the determinant operators are fully mixing with the single trace operator. 
On the other hand,  We are considering two open strings attached to the $D$-branes and a closed string. This problem is also realized by two determinants and a single trace operator in gauge theory side. However, we choose a simple setup where is no mixing between $D$-branes and strings. 
Namely, the single trace operator contracts with not full determinants but a small part of determinants.   
In other words,  we shall only study the contribution from interactions among strings and how to express the structure constant in terms of hexagon form factors.


\section{Set-up and Conventions}

We start by introducing set-up and conventions. 
Our main goal is to find a general expression for asymptotic three-point structure constants among two open strings and a closed string.
Concretely, the open string which we shall analyze is attached to the maximal giant graviton brane which is a spherical $D$-brane wrapping on $S^3$.
This is known as an integrability preserving open string configuration.\footnote{There are two different maximal giant gravitons which are respectively called $Y=0$ brane and $Z=0$ brane. Here we are using the embedding coordinates of $S^5$. Namely, $|X|^2+|Y|^2+|Z|^2 =1$. In the ungauged string sigma model, there is no difference among them since they are related to each other under $SO(6)$ rotation. However, if we choose a light-cone gauge direction and fix the gauge along the direction, the giant graviton brane splits into two groups. 
One includes the light-cone gauge direction and the other does not. The former is called $Y=0$ brane, and the latter is called $Z=0$ brane.
For example, the $Z=0$ brane is wrapping on the $S^3$ given as $X {\bar X}+Y{\bar Y} =1$. 
In the dual gauge theory, they are respectively realized by the following determinant operators: $\epsilon_{i_{1}\ldots i_{N}}^{j_{1}\ldots j_{N}} Z_{j_{1}}^{i_{1}}\ldots Z_{j_{N-1}}^{i_{N-1}} \left(Z^{L-M} \phi^{M} \right)_{j_{N}}^{i_{N}}$ and $\epsilon_{i_{1}\ldots i_{N}}^{j_{1}\ldots j_{N}} Y_{j_{1}}^{i_{1}}\ldots Y_{j_{N-1}}^{i_{N-1}} \left(Z^{L-M} \phi^{M} \right)_{j_{N}}^{i_{N}}$. In this paper, we shall only consider open strings attached to the $Y=0$ brane.} 
We would like to make the D-brane part and the open string part separate in computing a structure constant since the hexagon form factor is  originally designed by string worldsheet and is related  to  interacting strings. 
In other words, if there are nontrivial mixing between $D$-brane and string in the structure constant, there is no a priori reason that the structure constant can be expressed in terms of the hexagon form factors. 
Namely, we shall require some nice configurations where the $D$-brane only contracts with the other $D$-brane and the string only interacts with the other string if we would like to express the structure constant in terms of the hexagon.
This demand can be realized by choosing specialized set-ups and considering an only specific type of excitations.

We shall consider such set-ups for determinant operators and a single trace operator
There are two cases as follows: one is the three-point function of the following operators:
\begin{eqnarray}
{\cal O}_{1} &=& \epsilon_{i_{1}\ldots i_{N}}^{j_{1}\ldots j_{N}} Y_{j_{1}}^{i_{1}}\ldots Y_{j_{N-1}}^{i_{N-1}} \left(Z^{L_{1}} \right)_{j_{N}}^{i_{N}}, \qquad 
{\cal O}_{2} = \epsilon_{i_{1}\ldots i_{N}}^{j_{1}\ldots j_{N}} {{\bar Y}}_{j_{1}}^{i_{1}}\ldots {{\bar Y}}_{j_{N-1}}^{i_{N-1}} \left({\bar Z}^{L_{2}} \right)_{j_{N}}^{i_{N}}, \cr
{\cal O}_{3} &=& = \tr \left(\tilde{Z}^{L_3}\right) = \tr\left( \left(Z+{\bar Z}+X-{\bar X}\right)^{L_{3}}\right) ; \label{ooc}
\end{eqnarray}
the other is the problem of interactions among
\begin{eqnarray}
{\cal O}_1 &=& \tr(Z^{L_1}), \cr
{\cal O}_2 &=& \epsilon^{j_1 \dots j_N}_{i_1\dots i_N}Y^{i_1}_{j_1}\dots Y^{i_{N-1}}_{j_{N-1}}{(\bar{Z}^{L_2})}^{i_N}_{j_N},\quad
{\cal O}_3=\epsilon^{j_1\dots j_N}_{i_1\dots i_N}\bar{Y}^{i_1}_{j_1}\dots \bar{Y}^{i_{N-1}}_{j_{N-1}}{(\tilde{Z}^{L_3})}^{i_N}_{j_N}. \label{coo}
\end{eqnarray}
We can easily observe tha determinant operators are made of mostly $Y$-field or ${\bar Y}$-field. These correspond to the giant graviton brane part. With above choices, there is no contraction between the brane part and single trace operator.
The last pieces of the determinants would be open strings which can be contracted with the single trace operator.

Note that the above operators are reference states. We then compute three-point structure constants among operators obtained by putting magnonic excitations on top of the reference state.
Each operator ${\cal O}_3$ in both cases plays a role as a reservoir state even though they are different operators: one is the single trace and the other is the determinant. Therefore, we shall not put any excitations on ${\cal O}_3$.
We shall add magnons into ${\cal O}_1$ or ${\cal O}_2$ and also into both of ${\cal O}_1$ and ${\cal O}_2$ when we compute the tree level amplitudes. 
Furthermore, we have to choose well $SO(6)$ flavors of magnons not to interrupt our requirement.
For instance, we should not add a $Y$-excitation into ${\cal O}_1$ of the first class (\ref{ooc}).
This is because we would have nontrivial mixing between the $Y$-magnon in the open string part of ${\cal O}_1$ and a ${\bar Y}$ in $D$-brane part.


\section{Finite Coupling Conjecture}

We are interested in the three interacting strings vertex.
More precisely, we would like to express the structure constants for merging two open strings attached to the giant graviton branes into a closed string in terms of hexagon form factors, and give a reasonable conjecture for the structure constant at the finite coupling.
Holographically , we would like to study the structure constants of a single trace operator and two determinant operators in ${\cal N}=4$ SYM side.

\begin{figure}[t]
\begin{center}
\includegraphics[width=8cm]{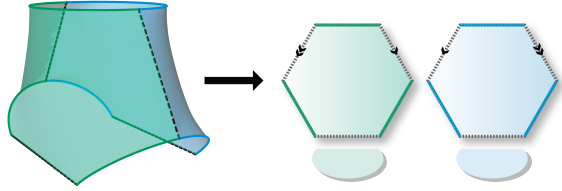}
\caption{Two hexagons are obtained by cutting the worldsheet. For going back to the original worldsheet diagram, we need to identify mirror edges together with gluing boundary states to two remainiing mirror edges.}
\label{hex}
\end{center}
\end{figure}

The key idea is to cut the worldsheet diagram given in Figure \ref{wsd} into two hexagon twist operators as in Figure \ref{hex}.
Then, each hexagon has three physical edges and three mirror edges where physical edges are the half of the closed string or the half of the open string.
Because two mirror edges at the bottom are surfaces of the open string end points, those should be related to the giant graviton branes.
The asymptotic expression where the lengths of string parts of gauge invariant operators are infinite simply reduces to the sum over partition related to distribution of magnons together with negative momenta contribution corresponding to reflections at boundaries.\footnote{Notice that there are two possible distributions to each hexagon even in open strings since we cut the open string itself to the half. This feature is quite new compared to \cite{Kim:2017phs} where open strings were not cut.}
On the other hand, if we would like to express a full finite coupling answer, one has to glue two hexagons back into the original worldsheet.
This gluing is to compute all finite size corrections which are to sum over all relevant mirror particles. 
Additionally, we should contract boundary states with the edges defined as endpoints of open strings where the boundary states can be defined as in \cite{Ghoshal:1993tm} through the exact reflection matrix \cite{Hofman:2007xp}.
 Inspiring from the intuition, we have a rough expression:
\begin{equation}
C_{123} \sim \int_{\rm mirrors} \int_{\rm boundaries} \left(\sum_{\rm magnons} {\cal H} \times {\cal H} \right),
\end{equation}
where ${\cal H}$ is the hexagon form factor, and the sum over magnons means the partition into two hexagons. On the other hand, two integrations over mirror particles and boundaries are related to finite size corrections which should be performed to obtain the original worldsheet by gluing edges and contracting with boundary states. 
Actually, the above expression is almost the same as the result in structure constants of single trace operators or closed strings. 
The difference is that the physical edges of the hexagon in our setup can be not only the half of the closed string but also the half of the open string. 
Two mirror edges from a hexagon should be identified with the other two mirror edges from the other hexagon for recovering two open strings and a closed string. 
This procedure would be realized by summing over mirror particles.  
Furthermore, we have to contract two boundary states for two remaining mirror edges from two hexagons. 
It makes sense to think such a contraction with boundary states as finite size corrections. 
Actually, when the volume is finite but still very large, mirror particle contributions would be reduced to the L\"ushcer corrections.
For closed string, the L\"ushcer corrections are related to the circulation of virtual particles along worldsheet cylinder.
Contracting the mirror edges at the bottom with the boundary states would be also related to finite size effects. 
Here, virtual particles can be created, propagate, reflect at each boundary and be annihilated back. These are known as the boundary   L\"ushcer corrections.
In this paper, we shall not analyze these finite size corrections. 
Instead of that, we would like to check the validity of our conjecture by directly computing tree-level structure constants at the weak coupling.

As mentioned before, the asymptotic structure constants should be simply read off from the hexagon form factors with summing over partitions and with momenta reflection effects. 
Let us give the all-loop expression for a structure constant in our setup. We consider the case (\ref{coo}) with $S_1$ magnons to ${\cal O}_1$ and $S_2$ magnons to ${\cal O}_2$.
We already know the corresponding expressions when $S_2$ is zero or $S_1$ is zero. Those are nothing but the asymptotically exact structure constants in pure closed and pure open spin chains. The explicit expression are available in (31) of \cite{BKV} and (116) of \cite{Kiryu:2018phb}.\footnote{We should use the exact reflection matrix preserving $SU(2|1)$ symmetry constructed in \cite{Hofman:2007xp} when we express the open-chain part. } 
Our case is a sort of mixed configuration between closed and open chains. Thus, we suggest the product of those as the asymptotically exact structure constant even though we shall not explicitly write down the product.
As in (31) of \cite{BKV} and (116) of \cite{Kiryu:2018phb}, it would be better to define the ratio between the structure constant of excited states with magnons and that of ground state without any mangon such as $\frac{C_{123}^{S_1 S_2 \circ}}{C_{123}^{\circ\circ\circ}}$ since we can then ignore unimportant combinatorial factors and the contraction between $D$-branes parts. On the other hand, the Gaudin norms from the closed chain and the open chain should be respectively included.
As the simplest example, let us consider the case for $S_1 =S_2 = 1$ at the weak coupling. Then, the norms are simply related to the lengths $L_{1,2}$. There would be no anyy $S$-matrix factor.
Lastly, one- particle hexagon form factors from each open and closed chains should be producted together.\footnote{As we shall comment later, there is additional sign changing.}


\section{Tree Level Analysis}
In this section, we compute tree level structure constants by inserting magnons into ${\cal O}_1$ or ${\cal O}_2$ in each three operators (\ref{ooc}) and (\ref{coo}). 
Since multi-magnon generalization is straightforward, we shall explicitly calculate the two-magnon problem.
In these computations, we shall use the following notations:\footnote{For further details, see Appendix A.}
\begin{itemize}
          \item$L_{1,2,3}$ are lenghs of $\mathcal{O}_{1,2,3}$, and $l_{jk}$ is the bridge length between $\mathcal{O}_{j}$ and $\mathcal{O}_{k}$.
          \item The numbering of spin-chain sites would be done in clockwise. On the other hand, the numbering of operators would be done in counter-clockwise.
    \item $p$ : magnon momentum in $\mathcal{O}_1$, \quad $q$ : magnon momentum in $\mathcal{O}_2$
    \item $e^{ip}\equiv\frac{u+\tfrac{i}{2}}{u-\tfrac{i}{2}}$, \,\ $e^{iq}\equiv\frac{v+\tfrac{i}{2}}{v-\tfrac{i}{2}}$ \,\ where $u$ and $v$ are rapidities for $p$ and $q$. 
    \item Reflection factor for $X$ or $\bar{X}$-magnon of the open spin-chain becomes  $R(p)=e^{ip}$. 
    \item One-magnon Bethe-Yang equation for closed chain is $e^{ip L}=1$, and that for open chain is $e^{2iq L}=1$ where the excitation is $X$ or $\bar{X}$.
          \item If we put two magnons on the $\mathcal{O}_1$, their momenta and rapidities will be denoted by $p_{1,2}$ and $u_{1,2}$.
\end{itemize}
We shall first suggest computation results in the next two subsections. After that, we shall summarize all the results together. We will interpret those physically compared to our conjecture.
Finally, some comments will be given.

\subsection{Open-Open-Closed}
We shall first consider the case (\ref{ooc}). Namely, we will not insert any magnon into the closed spin-chain. 
As we emphasized several times, we would like to choose some possible setup where any string part is not mixed with the $D$-brane parts. 
Furthermore, since the reflection factors for $X$ and $\bar{X}$ are the same each other, it is sufficient to consider only three cases of two-magnon: 
firstly $X$-excitation on $\mathcal{O}_1$ and $X$-excitation on $\mathcal{O}_2$,  secondly $X$-excitation on $\mathcal{O}_1$ and ${\bar X}$-excitation on $\mathcal{O}_2$, and
lastly   $XX$ excitations on $\mathcal{O}_1$. All the others can be obtained from these or should be excluded by our requirement.
The structure constants up to normalization can be simply computed through Wick contractions. If we further know correct eigenstates for spin-chains, the key computations are reduced to geometric series. A remark is that we need to use the Bethe equations for expressing the final answer through hexagon form factors.

\subsubsection{$\mathcal{O}_1 : X$, $\mathcal{O}_2 : X$}
In this case, the Wick contraction is indirect which means each magnon on $\mathcal{O}_1$ or $ \mathcal{O}_2$ is contracted only with the reservoir $\mathcal{O}_3$. Remember $\mathcal{O}_3$ is made of $\tilde{Z}=Z+\bar{Z}+X-\bar{X}$. Thus, one can obtain
\begin{alignat}{1}
    C_{123}&\propto\left( \sum\limits_{n=1}^{l_{13}} e^{ipn}+R(p)e^{-ipn} \right) \left( \sum\limits_{n=l_{12}+1}^{L_2} e^{iqn}+R(q)e^{-iqn}\right)\cr
    &\propto-e^{ipl_{13}}e^{iql_{12}}+e^{ipl_{13}}e^{-iql_{12}}+e^{-ipl_{13}}e^{iql_{12}}-e^{-ipl_{13}}e^{-iql_{12}}.\label{indirectNew}
\end{alignat}

\subsubsection{$\mathcal{O}_1 : X$, $\mathcal{O}_2 : \bar{X}$}
In this case, there are two possible Wick contractions. One is the indirect contraction through the reservoir we already treated. The other is the direct contraction between $X$ on  $\mathcal{O}_1$ and $\bar{X}$ on $\mathcal{O}_2$.  Summing those together, we get
\begin{alignat}{1}
    C_{123}&\propto\underbrace{\sum\limits_{n=1}^{l_{12}} \left(e^{ip(L_1-n+1)}+R(p)e^{-ip(L_1-n+1)}\right) \left(e^{iqn}+R(q)e^{-iqn}\right)}_{\text{direct}} \nonumber\\
    &\quad-\underbrace{\left( \sum\limits_{n=1}^{l_{13}} e^{ipn}+R(p)e^{-ipn} \right) \left( \sum\limits_{n=l_{12}+1}^{L_2} e^{iqn}+R(q)e^{-iqn}\right)}_{\text{indirect = \eqref{indirectNew}}}\cr
    &\propto\frac{u-v+i}{u-v}e^{ipl_{13}}e^{iql_{12}}-\frac{u+v+i}{u+v}e^{ipl_{13}}e^{-iql_{12}}\nonumber\\
    &\quad-\frac{u+v-i}{u+v}e^{-ipl_{13}}e^{iql_{12}}+\frac{u-v-i}{u-v}e^{-ipl_{13}}e^{-iql_{12}}
\end{alignat}

\subsubsection{$\mathcal{O}_1 : XX$}\label{XX}
Here,  as the open spin-chain for the $\mathcal{O}_1$  has two magnons, we should use the two-magnon wavefunction in Wick contraction. The Bethe ansatz is written through all the possible momenta combination including reflection effects. The result is given as
\begin{alignat}{1}
    C_{123}&\propto g(p_1,p_2)+R(p_1) g(-p_1,p_2)+S(p_1,p_2)S(-p_2,p_1)R(p_2)g(p_1,-p_2)\nonumber\\
    &\quad+S(p_1,p_2)S(-p_2,p_1)R(p_1)R(p_2)g(-p_1,-p_2),\cr
    &\propto\frac{u_1-u_2}{u_1-u_2+i}-S(p_1,p_2)S(p_1,-p_2)\frac{u_1+u_2}{u_1+u_2-i}e^{2ip_1l_{13}}\nonumber\\
    &\quad-\frac{u_1+u_2}{u_1+u_2+i}e^{2ip_2l_{13}}+S(p_1,p_2)S(p_1,-p_2)\frac{u_1-u_2}{u_1-u_2-i}e^{2i(p_1+p_2)l_{13}},
\end{alignat}
where we defined
\begin{alignat}{1}
    g(p_1,p_2)&\equiv\sum\limits_{1\leq n_1\leq n_2\leq L_1}e^{ip_1n_1}e^{ip_2n_2}+S(p_1,p_2)e^{ip_2n_1}e^{ip_1n_2}\cr
    &=\left(\frac{u_1-u_2}{u_1-u_2+i}-e^{ip_2l_{13}}-S(p_1,p_2)e^{ip_1l_{13}}+\frac{u_1-u_2}{u_1-u_2+i}e^{i(p_1+p_2)l_{13}}\right)\nonumber\\
    &\quad\times \frac{1}{i}(u_1+\tfrac{i}{2})\frac{1}{i}(u_2+\tfrac{i}{2}).
\end{alignat}

\subsection{Closed-Open-Open}

Let us move to the case (\ref{coo}). 
Now, the reservoir becomes $\mathcal{O}_3$  which is one of the determinants. Therefore, magnons can be inserted into the closed chain as well as the other open chain.
Most of the computations are parallel to those in the previous subsection except that the operator $\mathcal{O}_1$ is the trace operator so that Bethe eigenstate from $\mathcal{O}_1$ has no momentum reflection contribution.

\subsubsection{$\mathcal{O}_1 : X$, $\mathcal{O}_2 : X$}
As in the case (\ref{ooc}), the structure constant is calculated by indirect Wick contraction: 
\begin{alignat}{1}
    C_{123}&\propto\left( \sum\limits_{n=1}^{l_{13}} e^{ipn} \right)\left( \sum\limits_{n=l_{12}+1}^{L_2} e^{iqn}+R(q)e^{-iqn}\right)\cr
    &=\left( \dfrac{1}{i}(u+\tfrac{i}{2})\left(e^{ipl_{13}}-1\right)\right)\cr
    &\qquad\times\left( \dfrac{1}{i}(v+\tfrac{i}{2})\left(e^{iqL_2}-e^{iql_{12}}\right)-e^{iq}e^{-iq}\dfrac{1}{i}(v+\tfrac{i}{2})\left(e^{-iqL_2}-e^{-iql_{12}}\right)\right)\cr
    &\propto -e^{i(-p+q)l_{12}}+e^{i(-p-q)l_{12}}+e^{iql_{12}}-e^{-iql_{12}}.\label{indirect}
\end{alignat}

\subsubsection{$\mathcal{O}_1 : X$, $\mathcal{O}_2 : \bar{X}$}
The sum of direct and indirect contractions is written as
\begin{alignat}{1}
    C_{123}&\propto\underbrace{\sum\limits_{n=1}^{l_{12}} e^{ip(L_1-n+1)} \left(e^{iqn}+R[q]e^{-iqn}\right)}_{\text{direct}} -\underbrace{\left( \sum\limits_{n=1}^{l_{13}} e^{ipn} \right) \left( \sum\limits_{n=l_{12}+1}^{L_2} e^{iqn}+R[q]e^{-iqn}\right)}_{\text{indirect = \eqref{indirect}}}\cr
    &\propto\frac{u-v-i}{u-v}-\frac{u+v-i}{u+v}-e^{iql_{12}}+e^{-iql_{12}}\nonumber\\
    &\qquad+\frac{-u+v-i}{-u+v}e^{i(-p+q)l_{12}}-\frac{-u-v-i}{-u-v}e^{i(-p-q)l_{12}}.
\end{alignat}

\subsubsection{$\mathcal{O}_1 : XX$}
Finally, if we consider two-magnon in the closed chain, one can compute the structure constant such as 
\begin{alignat}{1}
    C_{123}&\propto\sum\limits_{1\leq n_1\leq n_2\leq L_1}e^{ip_1n_1}e^{ip_2n_2}+S(p_1,p_2)e^{ip_2n_1}e^{ip_1n_2}\nonumber\\
    &\equiv g(p_1,p_2)\nonumber\\
    &\propto\left(\frac{u_1-u_2}{u_1-u_2+i}-e^{ip_2l_{13}}-S(p_1,p_2)e^{ip_1l_{13}}+\frac{u_1-u_2}{u_1-u_2+i}e^{i(p_1+p_2)l_{13}}\right).
\end{alignat}

\subsection{Interpretation of Results}
We now would like to express the tree-level results in terms of the hexagon form factor and to interpret the results appropriately. 
The goal is to test our conjecture in section 3 at the weak coupling regime. 
Note that the underlined terms are surviving terms so that other terms cancel out themselves vertically. 
The hexagon form factors we shall use below  are as follows \cite{BKV}:
\begin{equation}
h_{X|X}(u,v)=\frac{u-v-i}{u-v},\quad h_{X|\bar{X}}=-1, h_X=-h_{\bar{X}}=1, \quad h_{XX}(u_1,u_2)=\frac{u_1-u_2}{u_1-u_2+i},
\end{equation}
which are the leading weak coupling expressions of the exact hexagon form factor constructed in \cite{BKV}.
The structure constants for two-magnon are summarized as below.
\subsubsection*{Open-Open-Closed}
\begin{alignat*}{1}
        &\mathcal{O}_1 : X,\quad \mathcal{O}_2 : X\\
                &\qquad C\propto +\left[h_{X|\bar{X}}(u,v)-h_X(v)h_{\bar{X}}(u)e^{ipl_{13}}-h_X(u)h_{\bar{X}}(v)e^{iql_{12}}+\underline{h_{X|\bar{X}}(v,u)e^{ipl_{13}}e^{iql_{12}}}\right]\\
                &\qquad\qquad-\left[h_{X|\bar{X}}(-u,v)-h_X(v)h_{\bar{X}}(-u)e^{-ipl_{13}}-h_X(-u)h_{\bar{X}}(v)e^{iql_{12}}+\underline{h_{X|\bar{X}}(v,-u)e^{-ipl_{13}}e^{iql_{12}}}\right]\\
                &\qquad\qquad-\left[h_{X|\bar{X}}(u,-v)-h_X(-v)h_{\bar{X}}(u)e^{ipl_{13}}-h_X(u)h_{\bar{X}}(-v)e^{-iql_{12}}+\underline{h_{X|\bar{X}}(-v,u)e^{ipl_{13}}e^{-iql_{12}}}\right]\\
                &\qquad\qquad+\left[h_{X|\bar{X}}(-u,-v)-h_X(-v)h_{\bar{X}}(-u)e^{-ipl_{13}}-h_X(-u)h_{\bar{X}}(-v)e^{-iql_{12}}+\underline{h_{X|\bar{X}}(-v,-u)e^{-ipl_{13}}e^{-iql_{12}}}\right]\\
                &\qquad\quad \propto h_{X|\bar{X}}(-u,-v)e^{2ipl_{13}}e^{2iql_{12}}-h_{X|\bar{X}}(u,-v)e^{2iql_{12}}-h_{X|\bar{X}}(-u,v)e^{2ipl_{13}}+h_{X|\bar{X}}(u,v)\\
        &\omit\vspace{0.2mm}\\
    &\mathcal{O}_1 : X,\quad \mathcal{O}_2 : \bar{X}\\
                &\qquad C\propto +\left[h_{X|X}(u,v)-h_X(v)h_X(u)e^{ipl_{13}}-h_X(u)h_X(v)e^{iql_{12}}+\underline{h_{X|X}(v,u)e^{ipl_{13}}e^{iql_{12}}}\right]\\
                &\qquad\qquad-\left[h_{X|X}(-u,v)-h_X(v)h_X(-u)e^{-ipl_{13}}-h_X(-u)h_X(v)e^{iql_{12}}+\underline{h_{X|X}(v,-u)e^{-ipl_{13}}e^{iql_{12}}}\right]\\
                &\qquad\qquad-\left[h_{X|X}(u,-v)-h_X(-v)h_X(u)e^{ipl_{13}}-h_X(u)h_X(-v)e^{-iql_{12}}+\underline{h_{X|X}(-v,u)e^{ipl_{13}}e^{-iql_{12}}}\right]\\
                &\qquad\qquad+\left[h_{X|X}(-u,-v)-h_X(-v)h_X(-u)e^{-ipl_{13}}-h_X(-u)h_X(-v)e^{-iql_{12}}+\underline{h_{X|X}(-v,-u)e^{-ipl_{13}}e^{-iql_{12}}}\right]\\
                &\qquad\quad \propto h_{X|X}(-u,-v)e^{2ipl_{13}}e^{2iql_{12}}-h_{X|X}(u,-v)e^{2iql_{12}}-h_{X|X}(-u,v)e^{2ipl_{13}}+h_{X|X}(u,v)\\
        &\omit\vspace{0.2mm}\\
        &\mathcal{O}_1 : XX\\
                &\qquad C\propto +\left[h_{XX}(u_1,u_2)-e^{ip_2l_{12}}-S(p_1,p_2)e^{ip_1l_{12}}+\underline{h_{XX}(u_1,u_2)e^{i(p_1+p_2)l_{12}}}\right]\\
                &\qquad\qquad-\left[h_{XX}(-u_1,u_2)-e^{ip_2l_{12}}-S(-p_1,p_2)e^{-ip_1l_{12}}+\underline{h_{XX}(-u_1,u_2)e^{i(-p_1+p_2)l_{12}}}\right]\\
                &\qquad\qquad-\left[h_{XX}(u_1,-u_2)-e^{-ip_2l_{12}}-S(p_1,-p_2)e^{ip_1l_{12}}+\underline{h_{XX}(u_1,-u_2)e^{i(p_1-p_2)l_{12}}}\right]S(p_1,p_2)S(-p_2,p_1)\\
                &\qquad\qquad+\left[h_{XX}(-u_1,-u_2)-e^{-ip_2l_{12}}-S(-p_1,-p_2)e^{-ip_1l_{12}} \right. \cr
&\qquad \qquad \qquad \left.   +\underline{h_{XX}(-u_1,-u_2)e^{i(-p_1-p_2)l_{12}}}\right]S(p_1,p_2)S(-p_2,p_1)\\
                &\qquad\quad\propto h_{XX}(u_1,u_2)-S(p_1,p_2)S(p_1,-p_2)h_{XX}(-u_1,u_2)e^{2ip_1l_{12}}-h_{XX}(u_1,-u_2)e^{2ip_2l_{12}}\\
                &\qquad\qquad+S(p_1,p_2)S(p_1,-p_2)h_{XX}(-u_1,-u_2)e^{2i(p_1+p_2)l_{12}}\\
        &\omit\vspace{0.2mm}\\
        \hline\\    
\end{alignat*}
\subsubsection*{Closed-Open-Open}
\begin{alignat*}{1}
        &\mathcal{O}_1 : X,\quad \mathcal{O}_2 : X \nonumber\\
                &\qquad C\propto \left[h_{X|\bar{X}}(u,v)-h_X(u)h_{\bar{X}}(v)e^{ipl_{13}}-\underline{h_X(u)h_{\bar{X}}(v)e^{iql_{12}}}+\underline{h_{X|\bar{X}}(v,u)e^{ipl_{13}}e^{iql_{12}}}\right]\\
                &\qquad\qquad-\left[h_{X|\bar{X}}(u,-v)-h_X(u)h_{\bar{X}}(-v)e^{ipl_{13}}-\underline{h_X(u)h_{\bar{X}}(-v)e^{-iql_{12}}}+\underline{h_{X|\bar{X}}(-v,u)e^{ipl_{13}}e^{-iql_{12}}}\right]\\
        &\omit\vspace{0.2mm}\\
        &\mathcal{O}_1 : X,\quad \mathcal{O}_2 : \bar{X}\\
                &\qquad C\propto \left[\underline{h_{X|X}(u,v)}-\underline{h_X(u)h_X(v)e^{iql_{12}}}-h_X(v)h_X(u)e^{ipl_{13}}+\underline{h_{X|X}(v,u)e^{ipl_{13}}e^{iql_{12}}}\right]\\
                &\qquad\qquad-\left[\underline{h_{X|X}(u,-v)}-\underline{h_X(u)h_X(-v)e^{-iql_{12}}}-h_X(-v)h_X(u)e^{ipl_{13}}+\underline{h_{X|X}(-v,u)e^{ipl_{13}}e^{-iql_{12}}}\right]\\
        &\omit\vspace{0.2mm}\\
        &\mathcal{O}_1 : XX\\
                &\qquad C\propto h_{XX}(u_1,u_2)-e^{ip_2l_{13}}-S(p_1,p_2)e^{ip_1l_{13}}+h_{XX}(u_1,u_2)e^{i(p_1+p_2)l_{13}}\\
\end{alignat*}
From these results, it is clear that the hexagonalization is working. We should consider all possible partitions of momenta into two hexagons together with appripriate propagation factors and $S$-matrix factors.
Furthermore, for open-chains, we should take into account the negative sign of momenta related to the reflection effects.
An important observation is that we should do the sign flipping operation when a magnon moves from a hexagon to the other hexagon, and when a magnon reflects at boundaries.
Although we could not nicely explain this observation, the same was noticed even in the hexagonliazaiton of structure constant problem among three single trace operators \cite{Les}.
All these features are nicely matched with our conjecture since the finite coupling expression is asymptotically reduced to the partition sums of momenta and the sums over all possible sign changes of momenta at the weak coupling.

\subsection{Comment on Norm}
In our setup, if we would like to express $C_{123}$ completely, the expression should include the norm ${\mathbb N}$ which is given by
\begin{equation}
{\mathbb N} = {\cal N}_{open} \times {\cal N}_{closed} \times {\cal M}. 
\end{equation}
Here, ${\cal N}_{open}$ and ${\cal N}_{closed}$ are respectively given by Gaudin formulas which can be computed  from open and closed string Bethe-Yang equations.\footnote{For the open-chain from Wilson loop, the norm formula was studied in \cite{Kim:2017phs}.} 
On the other hand,  ${\cal M}$ is a normalization factor coming from contractions between $D$-brane parts.
Surely, since we use the ratio of $C_{123}^{S_{1} S_{2}\circ}$ and $ C_{123}^{\circ\circ\circ}$ in our conjecture, we did not need to exxplictly know the expression ${\cal M}$.
Nevertheless, by computing two-point functions of determinants corresponding to $D$-branes, one could determine ${\cal M}$ even though we shall not perform this task.\footnote{See \cite{Bissi:2011dc, Caputa:2012yj, Jiang:2019xdz} for further discussion.}
Note that one of the most special aspects in our setup was what one can separately compute ${\cal M}$ regardless of configurations of open strings.

\subsection{Multi-particle generalization}
In this paper, we have explicitly computed the structure constants of operators with two-particle excitations.  However, the multi-particle generalization is straightforward and we could check our conjecture is indeed working beyond two-magnon. 
For example, let us introduce a possible three-magnon computation where two magnons are inserted into the closed chain and a magnon is putted on one of the open chains.. 
We then can obtain
\begin{eqnarray}
	C_{123} &\propto& \left(\sum\limits_{1\leq x_1\leq x_2\leq L_1}e^{ip_1x_1}e^{ip_2x_2}+S(p_1,p_2)e^{ip_2x_1}e^{ip_1x_2}\right) \left( \sum\limits_{n=l_{12}+1}^{L_2} e^{iqn}+R[q]e^{-iqn}\right)\cr
				&=& h_{XX|\bar{X}}(u_1,u_2,v)\underline{-h_{XX}(u_1,u_2)h_{\bar{X}}(v)e^{iql_{13}}}-h_{X|\bar{X}(u_1,v)h_X(u_2)e^{ip_1l_{13}}}\cr
				& &-h_{X|\bar{X}}(u_2,v)h_X(u_1)e^{ip_2l_{13}}+h_{\bar{X}}(v)h_{XX}(u_1,u_2)e^{i(p_1+p_2)l_{13}}+\underline{h_{X}(u_1)h_{X|\bar{X}}(u_2,v)e^{i(p_2+q)l_{13}}}\cr
				& &+\underline{h_X(u_2)h_{X|\bar{X}}(u_1,v)S(p_1,p_2)e^{i(p_1+q)l_{13}}}-\underline{h_{XX|\bar{X}}(u_1,u_2,v)e^{i(p_1+p_2+q)l_{13}}}-\left( q\leftrightarrow-q \right)\cr
				&=& \left[h_{XX}(u_1,u_2)e^{iql_{13}}-e^{i(p_2+q)l_{13}}-S(p_1,p_2)e^{i(p_1+q)l_{13}}+h_{XX}(u_1,u_2)e^{i(p_1+p_2+q)l_{13}}\right]-\left( q\leftrightarrow-q \right),\nonumber
\end{eqnarray}
where the underlined terms are surviving terms as before.
This result is again matched well with our conjecture based on the hexagonalization.

\section{Discussion}

In this paper, we studied the interaction among a closed string and two open strings attached to the giant graviton based on integrability point of view. 
We proposed how to express structure constants of a single trace operator and two determinant operators at finite coupling, and checked its validity at the weak coupling by directly computing tree level structure constants.
Our approach was to consider recombination of string worldsheet through fundamental building blocks which are hexagon form factors. 
Since the finite coupling conjecture is consistent with perturbative results, it seems that the hexagonalization to the three-string worldsheet is universal.
It would be interesting to see if the hexagonalization can be applied again to the case where two closed strings and an open string are interacting.

It would be also interesting to study four-point functions for single trace operators and determinant operators. 
The simplest set-up would be the interaction among two closed and two open strings since there exists at least a setup where the $D$-branes part can be decoupled from the string part as in this paper.
For example, we may consider
\begin{eqnarray}
{\cal O}_{1} &=& \epsilon_{i_{1}\ldots i_{N}}^{j_{1}\ldots j_{N}} Y_{j_{1}}^{i_{1}}\ldots Y_{j_{N-1}}^{i_{N-1}} \left(X^{K} \right)_{j_{N}}^{i_{N}}, \qquad\; 
 {\cal O}_{2} = \tr \left( Z^{\frac{K}{2}} {\bar X}^{\frac{K}{2}} \right)  + \ldots , \cr
{\cal O}_{3} &=& \epsilon_{i_{1}\ldots i_{N}}^{j_{1}\ldots j_{N}} {{\bar Y}}_{j_{1}}^{i_{1}}\ldots {{\bar Y}}_{j_{N-1}}^{i_{N-1}} \left({\bar Z}^{K} \right)_{j_{N}}^{i_{N}}, , \qquad {\cal O}_{4} = \tr \left( Z^{\frac{K}{2}} {\bar X}^{\frac{K}{2}} \right) + \ldots . \nonumber 
\end{eqnarray}
Then, it would be also wonderful to relate the result for four-point functions based on hexagon in large charge limit with a result based on octagon \cite{Bargheer:2019kxb}.

In this paper, we have never studied both loop corrections and finite size corrections. If our conjecture makes sense indeed, it should be available to calculate those.
For example, applying techniques in \cite{Kiryu:2018phb} and \cite{Basso:2015eqa} to our setup might be doable. In particular,  it would be interesting to study the leading finite size correction related to the boundary states in the philosophy of \cite{Bajnok:2010ui} for the spectral problem. 

Finally, studying the non-planar corrections would be one of the most interesting future directions. 
There are two nonplanar situations. One is simply coming from the finite $N$. Even in this finite $N$ physics, the hexagonalization idea seems to work \cite{Bargheer:2017nne, Bargheer:2018jvq}. Thus, it would be interesting to calculate the leading finite $N$ correction in our setup.
The other nonplanarity is related to heavy operators in the ${\cal N}=4$ SYM. If such operators scale as ${\cal O}(N^2)$ in the large $N$, the dual gravity is backreacted \cite{llm}.
Remarkably, even though the string on the LLM geometry is in general non-integrable, integrable subsectors which are realized from the condition of planar excitations and specific boundary conditions were proposed \cite{llmmagnon, deMelloKoch:2018tlb}, and proved \cite{deMelloKoch:2018ert, kochkim}. 
It would be nice to study if the hexagonalization can be applied to the integrable subsectors \cite{kimschur}.


{\vskip 0.25cm}
\noindent
\begin{centerline} 
{\bf Acknowledgements}
\end{centerline} 

The work of KK and KL is supported by Basic Science Research Program through the National Research Foundation of Korea(NRF) funded by the Ministry
of Science, ICT \& Future Planning(NRF- 2017R1A2B4004810) and GIST Research Institute(GRI) grant funded by the GIST in 2019, whereas the research of MK is supported by the South African Research Chairs Initiative of the Department of Science and Technology and National Research Foundation as well as funds received from the National Institute for Theoretical Physics (NITheP).

\appendix

\section{Review on Open Spin-Chain}
Here we review the open spin-chain which is holographically dual to the open string attached to the $Y=0$ brane \cite{Berenstein:2005vf}.
The Hamiltonian is constructed by evaluating relevant boundary Feynman diagrams as well as bulk Feynman diagrams \cite{Berenstein:2005vf} and is explicitly given as
\begin{eqnarray}
        \mathbb{H}&=&\frac{1}{2}\lambda\sum_{j=2}^{L-2}[ {\mathbb K}_{j,j+1}+2({\mathbb I}_{j,j+1}- {\mathbb P}_{j,j+1})]+ \frac{1}{2} \lambda {\mathbb Q}_1^Y [ {\mathbb K}_{1,2}+2( {\mathbb I}_{1,2}- {\mathbb P}_{1,2})] {\mathbb Q}_1^Y\cr
        &+& \frac{1}{2}\lambda {\mathbb Q}_L^Y [ {\mathbb K}_{L-1,L}+2( {\mathbb I}_{L-1,L}- {\mathbb P}_{L-1,L})] {\mathbb Q}_L^Y +\lambda( {\mathbb I}- {\mathbb Q}_1^{\bar{Y}})+\lambda( {\mathbb I}-{\mathbb Q}_L^{\bar{Y}}),
\end{eqnarray}
where ${\mathbb I}_{j,j+1}$, ${\mathbb P}_{j,j+1}$ and ${\mathbb K}_{j,j+1}$ are respectively the identity operator, the permutation operator and the trace operator acting on two adjacent sites $(j, j+1)$.
The operator $Q^\phi_n$ is a projection operator defined as
\begin{eqnarray}
&& Q^\phi_n \mid \cdots \stackrel{\stackrel{n}{\downarrow}}{\phi} \cdots \rangle= 0 \cr
&& Q^\phi_n \mid \cdots \stackrel{\stackrel{n}{\downarrow}}{\psi} \cdots \rangle= \mid \cdots \stackrel{\stackrel{n}{\downarrow}}{\psi} \cdots \rangle. \nonumber
\end{eqnarray}
Namely, if the field at the $n$-th site is the same with $\phi$ in $Q^\phi_n$, the value vanishes. On the other hand, if the field is different, $Q^\phi_n$ just becomes the identity operator.
One can notice that the projection operators are only related to $Y$-field or ${\bar Y}$-field.

Since the Hamiltonian is commute with the total spin number operator, one can classify the whole Hilbert space in terms of numbers of excitations: ${\cal H}= {\cal H}_{1} \oplus {\cal H}_{2} \oplus {\cal H}_{3} \oplus \cdots \oplus {\cal H}_{L/2}$.
Furthermore, we shall assume that the eigenstates can be written through the sum of plane waves with well-defined lattice momenta and their reflections with a negative sign of momenta. For example, the one-magnon eigenstate becomes
\begin{equation}
        \mid \psi \rangle=\sum^{L_{\rm eff}}_{n=1_{\rm eff}} (e^{ipn}+R(p)e^{-ipn})\mid n \rangle
\end{equation}
where we used $1_{\rm eff}$ and $L_{\rm eff}$ respectively as the first site and the last site of the spin-chain. 
This is because those can be differently defined for the type of excitations. It is not strange if we remind that the $Y$-excitation cannot be located at the first site. 

Since $Z$ and ${\bar Z}$ can be thought as sort of bound states, there would be in general four different scalar excitations: $X$, ${\bar X}$, $Y$ and ${\bar Y}$.
However, we shall further exclude $Y$-excitation and ${\bar Y}$-excitation from our consideration because those make interactions between $D$-brane parts of the determinants and strings nontrivial. For other scalar excitations, see the original paper \cite{Berenstein:2005vf}.
For example, if we consider the following situation
\begin{eqnarray}
{\cal O}_{1} &=& \epsilon_{i_{1}\ldots i_{N}}^{j_{1}\ldots j_{N}} Y_{j_{1}}^{i_{1}}\ldots Y_{j_{N-1}}^{i_{N-1}} \left(Z^{L_{1}-1} Y \right)_{j_{N}}^{i_{N}}, \qquad 
{\cal O}_{2} = \epsilon_{i_{1}\ldots i_{N}}^{j_{1}\ldots j_{N}} {{\bar Y}}_{j_{1}}^{i_{1}}\ldots {{\bar Y}}_{j_{N-1}}^{i_{N-1}} \left({\bar Z}^{L_{2}-1 }{\bar Y} \right)_{j_{N}}^{i_{N}}, \cr
{\cal O}_{3} &=& = \tr \left(\tilde{Z}^{L_3}\right) = \tr\left( \left(Z+{\bar Z}+X-{\bar X}\right)^{L_{3}}\right) 
\end{eqnarray}
the excitations $Y$ and ${\bar Y}$ shall be contracted with ${\bar Y}$ and $Y$ in the $D$-brane part.
Thus, in our problem, it would be enough if we know the coordinate Bethe ansatz for $X$- and ${\bar X}$-excitations and the corresponding reflection factor.

Let us start with one magnon problem. 
Note that there is actually no difference between  $X$- and ${\bar X}$-excitations since they are symmetric in $Z$-vacuum spin-chain of $Y=0$ brane.
Thus, without loss of generality, we consider only $X$-excitation. For convenience, we further choose $\lambda=1$.
We then have the much simpler Hamiltonian
\begin{equation}
    \mathbb{H}=\sum_{l=1}^{L-1} (I_{l,l+1}-P_{l,l+1}). \nonumber
\end{equation}
If we assume the following one-magnon Bethe ansatz:
\begin{equation}
        \mid \psi \rangle=\sum^L_{n=1} (e^{ipn}+R(p) e^{-ipn})\mid n \rangle,
\end{equation}
it is straightforward to check that the Hamiltonian can be diagonalized such as 
\begin{eqnarray}
        \mathbb{H} \mid \psi \rangle &=& [(e^{ip}+R(p)e^{-ip})-(e^{2ip}+R(p)e^{-2ip})] \mid 1 \rangle +[2(e^{2ip}+R(p)e^{-2ip})-(e^{ip}+R(p)e^{-ip})-(e^{3ip}+R(p)e^{-3ip})]\mid 2 \rangle  + \cdots \cr
                                 &+& [2(e^{3ip}+R(p)e^{-3ip})-(e^{2ip}+R(p)e^{-2ip})-(e^{4ip}+R(p)e^{-4ip})]\mid 3 \rangle+[(e^{ipL}+R(p)e^{-ipL})-(e^{ip(L-1)}+R(p)e^{-ip(L-1)})]\mid L \rangle \cr
                                 &=& E\mid \psi \rangle \nonumber
\end{eqnarray}
if we demand the following identifications: 
\begin{eqnarray}
        && [(e^{ip}+R(p)e^{-ip})-(e^{2ip}+R(p)e^{-2ip})]\mid 1\rangle=E(e^{ip}+R(p)e^{-ip})\mid 1 \rangle,\\
        && [2(e^{2ip}+R(p)e^{-2ip})-(e^{ip}+R(p)e^{-ip})-(e^{3ip}+R(p)e^{-3ip})]\mid 2 \rangle=E(e^{2ip}+R(p)e^{-2ip})\mid 2 \rangle,\\
        && \qquad\vdots\\
        && [(e^{ipL}+R(p)e^{-ipL})-(e^{ip(L-1)}+R(p)e^{-ip(L-1)})]\mid L \rangle=E(e^{ipL}+R(p)e^{-ipL})\mid  L \rangle.
    \end{eqnarray}
One can easily compute the Bethe equation, the reflectionn matrix and the energy eigenvaluefrom these equations. 
Let us summarize the results:
\begin{equation}
e^{2ipL}=1, \quad R(p)=e^{ip} ,\quad E=4\sin^2 \left(\frac{p}{2} \right).
\end{equation}

The two-magnon problem can be similarly studied. Firstly, we need to remember there are only two cases: $YY$ and ${\bar Y} {\bar Y}$.
This is because $Y {\bar Y}$ can generate all other $SO(6)$ ingredients such as $X{\bar X}$ and $Z{\bar Z}$.
 One can start with the wavefunction such as
\begin{equation}
|\psi \rangle = \sum_{n_{1} < n_{2} } f(n_1, n_2 ) |n_1 ,n_2 \rangle ,\quad 
|n_1, n_2 \rangle=|Z \cdots Z \stackrel{\stackrel{n_{1}}{\downarrow}}{X} Z \cdots Z \stackrel{\stackrel{n_{2}}{\downarrow}}{X} Z \cdots Z \rangle.
\end{equation}
We then need to solve $H |\psi \rangle = E |\psi \rangle$ with a general ansatz $f(n_1 , n_2)$ such as
\begin{equation}
f(x_1 ,x_2 )= A_{1} e^{i p_1 n_1 + i p_2 n_2 } + A_{2} e^{-i p_1 n_1 + i p_2  n_2 } + \cdots A_{8} e^{-i p_2 n_1 - i p_1 n_2 }. 
\end{equation}
It turned out that the correct wavefunction is
\begin{eqnarray}
f(x_1 ,x_2 )&=&  e^{i p_1 n_1 + i p_2 n_2 } + R(p_{1})e^{-i p_1 n_1 + i p_2  n_2 } +R(p_{2})S(p_2 ,p_1 ) S(p_1 , -p_2 )e^{i p_1 n_1 - i p_2 n_2 } \cr
&+& R(p_{1})R(p_{2}) S(p_2 ,p_1 ) S(p_1 , -p_2 ) e^{-i p_1 n_1 - i p_2 n_2 } + S(p_2 ,p_1 ) e^{i p_2 n_1 + i p_1 n_2 } \cr
&+&R(p_{1}) S(p_1 , -p_2 ) e^{i p_2 n_1 - i p_1 n_2 } + R(p_{2}) S(p_2 ,p_1 ) e^{-i p_2 n_1 + i p_1 n_2 } \cr
&+& R(p_1 )R(p_2 ) S(p_1 , -p_2 )e^{-i p_2 n_1 - i p_1 n_2 }  
\end{eqnarray}
where the boundary amplitude $R(p)$ and the bulk scattering amplitude $S(p_1 ,p_2 )$ are given as
\begin{equation}
R(p) = e^{i p} , \qquad S(p_1 ,p_2 )= -\frac{1-2e^{i p_{1}}+e^{i p_{1} + i p_{2}}}{1-2e^{i p_{2}}+e^{i p_{1} + i p_{2}}} .  
\end{equation}
With the above wavefunction, we can really check that we solve the eigenvalue equation.
The eigenvalue $E$ is obtained as
\begin{equation}
E = 4 - e^{i p_{1}}- e^{-i p_{1}}- e^{i p_{2}}- e^{-i p_{2}} = 4\sin^2 \left(\frac{p_{1}}{2} \right) + 4\sin^2 \left(\frac{p_{2}}{2} \right).
\end{equation}
Moreover, for consistency, the following BAEs should be satisfied :
\begin{eqnarray}
e^{2 i p_{1} L} &=& S(p_{1},p_{2}) R(p_1) S(p_{1},-p_{2}) R(-p_1 )\\
e^{2 i p_{2} L} &=& S(p_{2},p_{1}) R(p_2) S(p_{2},-p_{1}) R(-p_2 ).
\end{eqnarray}
All results for two-magnon are consistent with one-particle reduction. 
Although we shall not explicitly write down, the multi-magnon generalization can be straightforwardly performed.

\vfill\eject
\end{document}